\documentclass[conference]{IEEEtran}
\IEEEoverridecommandlockouts
\usepackage{Shortcuts}
\def\BibTeX{{\rm B\kern-.05em{\sc i\kern-.025em b}\kern-.08em
    T\kern-.1667em\lower.7ex\hbox{E}\kern-.125emX}}
\begin{document}

\title{The Penalty in Scaling Exponent for Polar Codes\\is Analytically Approximated by the Golden Ratio
}

\author{\IEEEauthorblockN{Ori Shental}
\IEEEauthorblockA{\textit{Communications Theory Department} \\
\textit{Bell Labs}\\
Holmdel, New Jersey 07733, USA \\
ori.shental@nokia-bell-labs.com}
}

\maketitle

\begin{abstract}
The polarization process of conventional polar codes in binary erasure channel (BEC) is recast to the Domany-Kinzel cellular automaton model of directed percolation in a tilted square lattice. Consequently, the former's scaling exponent, $\mu$, can be analogously expressed as the inverse of the percolation critical exponent, $\beta$. Relying on the vast percolation theory literature and the best known numerical estimate for $\beta$, the scaling exponent can be easily estimated as $\mu_{\text{num}}^{\text{perc}}\simeq1/0.276486(8)\simeq3.617$, which is only about $0.25\%$ away from the known exponent computation from coding theory literature based on numerical approximation, $\mu_{\text{num}}\simeq3.627$. Remarkably, this numerical result for the critical exponent, $\beta$, can be analytically approximated (within only $0.028\%$) leading to the closed-form expression for the scaling exponent $\mu\simeq2+\varphi=2+1.618\ldots\simeq3.618$, where $\varphi\triangleq(1+\sqrt{5})/2$ is the ubiquitous golden ratio. As the ultimate achievable scaling exponent is quadratic, this implies that the penalty for polar codes in BEC, in terms of the scaling exponent, can be very well estimated by the golden ratio, $\varphi$, itself.
\end{abstract}

%\begin{IEEEkeywords}
%component, formatting, style, styling, insert
%\end{IEEEkeywords}

\section{Introduction}
Polar codes are capacity-achieving codes with an unprecedented diagrammatic construction~\cite{Arikan2009}. The latter property, along with a relatively simple decoding procedure, make polar codes a popular and attractive candidate for error correction in the design of state-of-the-art communication systems. For example, polar codes were recently chosen to encode the transmission of the control channels in the fifth generation (5G) cellular standardization.

One important metric of any error-correcting code (ECC) is its \emph{scaling exponent}. The scaling exponent, $\mu$, quantifies, per a given target decoding error probability, how fast a rate-$R$ code’s blocklength, $N$, increases as the gap to channel capacity, $C$, diminishes, or namely
\BE\label{eq_scaling_exponent}
    N\propto(C-R)^{-\mu}.
\EE

While the ultimate scaling exponent, achieved by random binary coding~\cite{Strassen1962}, is quadratic, conventional polar codes (\ie, using Arikan’s original $2\times2$ binary kernel~\cite{Arikan2009}) approach capacity rather slowly\footnote{A recent discussion on polar code constructions based on $l\times l$ binary polarization kernels with lower scaling exponents can be found in~\cite{Yao2019}.}. For instance, their scaling exponent for the binary erasure channel (BEC) was \emph{numerically} computed, based on a heuristic eigen-analysis of the polarization operator, to be about $\mu_{\text{num}}\simeq3.627$~\cite{Hassani2014}. For such a case, lower and upper analytical bounds on the scaling exponent were derived, originally in~\cite{Hassani2014} with the upper bound being further improved in~\cite{Mondelli2016}, to yield $3.579\leq\mu\leq3.639$ (The upper bound for any binary-input memoryless output-symmetric channel is $4.714$, and it was also conjectured that the lower bound can get tighter up to the BEC's scaling exponent~\cite{Hassani2014}.) In~\cite{Hassani2014}, also another lower bound, of closed-form nature, was suggested to give $\mu\geq(1-\frac{1}{2\log{2}})^{-1}\simeq3.589$. However, to date an exact explicit closed-form analytical expression for the scaling exponent remains unknown, even for the simple case of transmission over the BEC.

\begin{figure}[t!]
    \centering
    \includegraphics[width=\columnwidth,bb=0 0 307 163]{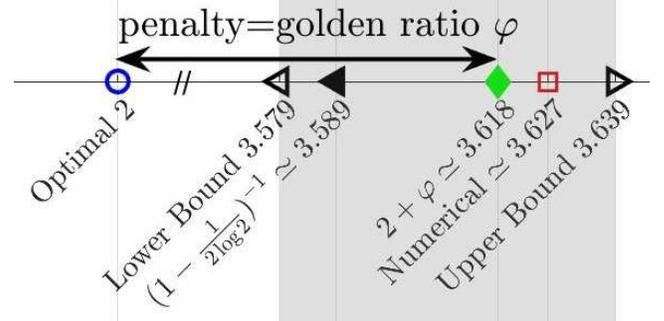} 
    \caption{The scaling exponent, $\mu$, of conventional polar code in BEC. The marker {\color{green} $\blacklozenge$} denotes this contribution's newly derived analytical approximation, being exactly $\varphi$ above the optimal value ({\color{blue} $\circ$}). Also marked for comparison, previous heuristic numerical computation ({\color{red}$\square$},~\cite{Hassani2014}) and known bounds ({\color{black}$\vartriangleleft,\blacktriangleleft$},\cite{Hassani2014} and {\color{black}$\vartriangleright$},\cite{Mondelli2016}).}
    \label{fig_results}
\end{figure}

In this contribution, an analytical closed-form accurate approximation of the scaling exponent of conventional polar codes in BEC is explicitly derived to interestingly yield
\BE\label{eq_main_result}
    \mu\simeq2+\varphi\simeq3.618,
\EE
where
\BE
    \varphi\triangleq\frac{(1+\sqrt{5})}{2}\simeq1.618
\EE
is the legendary \emph{golden ratio}. Fig.~\ref{fig_results} depicts the derived approximation~\eqref{eq_main_result} w.r.t. the optimal scaling exponent and aforementioned numerical and bounding results. This closed-form approximation is shown to be quite precise, beautifully falling within the relatively narrow allowable range as defined by the known bounds, and is only about a quarter percent away from the heuristic numerical estimate. Furthermore, based on~\eqref{eq_main_result}, the main takeaway of this contribution is that the penalty, or excess, in scaling exponent for conventional polar codes in BEC, w.r.t. the best possible scaling exponent of $2$, is very well approximated by the golden number $\varphi$.

The golden ratio, which goes by many names including the \emph{divine proportion}, has appeared ubiquitously across the history of science and art. It dates back to Euclid's "Elements" around 300 B.C., 1509's "De Divina Proportione" written by Luca Pacioli and illustrated by Leonardo da Vinci, Johannes Kepler's triangle (and limit of the ratios of successive terms of the Fibonacci sequence) around 1600 and even in Dan Brown's 2003 best-selling novel, “The Da Vinci Code”. More on the fascinating history of this mythical number can be found, for instance, in~\cite{Livio2018,Meisner2018}. It is quite remarkable how this fundamental number finds its way into polar coding and can be intimately connected into the intrinsic properties of such a modern artifact.

The proof of the proposed approximation stems from identifying a strong resemblance between the notion of channel polarization from coding theory, which is the `magic sauce' behind the workings of polar codes, and the statistical physics' concepts of Domany-Kinzel (DK) cellular automaton and directed percolation (DP) in a two-dimensional grid~\cite{Domany1984,Kinzel1985,Henkel2008}. Percolation theory, in general, describes the behavior of connected clusters in large random graphs \cite{Stauffer2014}. One of its evident manifestations is coffee percolation, modeling the way water filters through the ground beans to deliver the essential coffee brew. In DP the water is restricted to sink in only along a certain direction in space. To the best of our knowledge, this is the first time the DK model of cellular automaton, DP theory and their universal scaling laws are applied as tools for analyzing modern ECCs.

\section{Problem Formulation}
Consider the transmission over a $\text{BEC}(z)$ with erasure parameter, $z\in(0,1)$, of a binary polar code of block length $N=2^{n}$ generated by Arikan's $2\times2$ kernel, where $n\in\mathbb{Z}^{\ast}$ is the number of channel polarization stages used in the code's construction~\cite{Arikan2009}. Under successive cancellation decoding, the synthetic channel's erasure parameter polarizes (almost surely) to $\{0,1\}$-valued random variable $Z_{\infty}$. Furthermore, the evolution of this polarizing erasure parameter, $Z_{n}$, is known to be determined by the following closed-form recursion\footnote{More about the intricacies of polar codes can be found, for instance, in~\cite{Arikan2009,Hassani2014,Mondelli2016}.}
\begin{equation}\label{eq_Z_rec}
  Z_{n+1}=\begin{cases}
    Z_{n}^{2} , & \text{ w.p. $\frac{1}{2}$},\\
    1-(1-Z_{n})^{2}, & \text{ w.p. $\frac{1}{2}$},
  \end{cases}
\end{equation}
where $Z_{0}=z$.

Let $0<c<d<1$ be real constants and consider
\BE
    p_{n}(z,c,d)\triangleq\Pr(Z_{n}\in[c,d]|Z_{0}=z),
\EE the probability that the erasure parameter at polarizing stage $n$ in a $\text{BEC}(z)$ gets a value within the range $[c,d]$. Following~\eqref{eq_Z_rec}, this probability satisfies the recursion rule
\BE\label{eq_p_rec}
    p_{n+1}(z,c,d)=\frac{p_{n}(z^{2},c,d)+p_{n}(2z-z^{2},c,d)}{2},
\EE
with $p_{0}(z,c,d)=\mathds{1}_{z\in[c,d]}$.

Now, we seek the limiting rate, $\tilde{\mu}^{-1}\in(0,\infty)$, in which the probability of having an unpolarized channel (exponentially) decays to zero, or namely
\BE\label{eq_limiting_value}
    \tilde{\mu}^{-1}\triangleq-\lim_{n\rightarrow\infty}\frac{1}{n}\log_{2}\big(p_{n}(z,c,d)\big).
\EE
A proof that $\tilde{\mu}=\mu$, \ie\, the reciprocal of the limiting value of~\eqref{eq_limiting_value} equals the scaling exponent, as defined in~\eqref{eq_scaling_exponent}, can be found in~\cite{Hassani2014}.
In the following section the main result, as stated in~\eqref{eq_main_result}, is derived.

\section{Analytical Approximation}
Evidently due to polarization, the decay of the probability $p_{n}(z,c,d)$ is driven by its limiting behavior in the two extreme regimes as $Z_{n}$ symmetrically approaches either the noiseless $0$ or useless $1$ erasure values. Hence, it is sufficient to study and evaluate the decay of $p_{n}(z,c,d)$, according to \eqref{eq_p_rec}, in the regime of $Z_{n}\rightarrow0$.

\begin{figure}[t!]
    \centering
    \includegraphics[width=\columnwidth,bb=0 0 225 172]{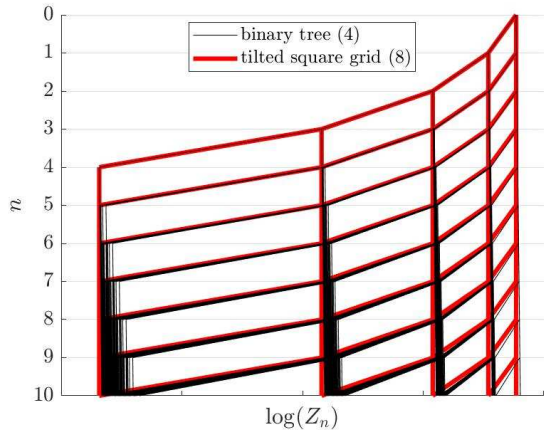}
    \caption{Binary tree representing~\eqref{eq_Z_rec} and a tilted square grid corresponding to~\eqref{eq_Z_rec_approx}. For a small $Z_{0}=z$ initialization, the square grid very well approximates the polarization process' binary tree.}
    \label{fig_tree_grid}
\end{figure}

Based on~\eqref{eq_Z_rec}, the random process, $Z_{n}$, underlying the channel polarization mechanism exhibited by the polar code in the BEC, can be clearly represented by a (perfect) \emph{binary tree} with a root and $n$ layers, each with $2^{n}$ nodes. However, this polarization process' binary tree, ordinary at first sight, has a peculiar, yet as we shall see very useful, attribute: As $Z_{n}$ diminishes, certain adjacent nodes in each layer of the tree become `sticky' and (almost) indistinguishable, and the tree (which is by definition an infinite-dimensional lattice) resembles a \emph{two-dimensional (tilted) square lattice}. This behavior is illustrated in Fig.~\ref{fig_tree_grid} plotting the evolution of the logarithm of $Z_{n}$ across the polarization stages $n$.

This attribute can be explained by the fact that in the small $Z_{n}$ regime, the leading quadratic and linear orders of $Z_{n}^{2}$ and $2Z_{n}-Z_{n}^{2}$, respectively, dominate, and the recursion rule~\eqref{eq_Z_rec} can be very well approximated by the simplified recursion rule
\begin{equation}\label{eq_Z_rec_approx}
  Z_{n+1}=\begin{cases}
    Z_{n}^{2} , & \text{ w.p. $\frac{1}{2}$},\\
    Z_{n}, & \text{ w.p. $\frac{1}{2}$},
  \end{cases}
\end{equation}
where $Z_{0}$ equals (some small) $z$.\footnote{Clearly the trivial recursion $Z_{n}\rightarrow Z_{n+1}$ does not support the existence of the complementary polarization towards $1$. This subtlety is addressed and circumvented in Section~\ref{sec_DK}.\label{foonote_trivial}}

Consequently, the recursive tree update rule~\eqref{eq_p_rec} of the probability of having an unpolarized
channel, $p_{n}(z,c,d)$, can be approximated via a random process on a square grid with
\BE\label{eq_p_rec_approx}
    p_{n+1}(z,c,d)\simeq\frac{p_{n}(z^{2},c,d)+p_{n}(z,c,d)}{2}.
\EE
This important observation allows recasting the original polarization problem on a binary tree, from the domain of coding theory, into a DP analysis in a square grid, well-known as the DK cellular automaton.

\subsection{Domany-Kinzel Cellular Automaton}\label{sec_DK}

\begin{figure}[t!]
    \centering
    \includegraphics[width=\columnwidth,bb=0 0 225 172]{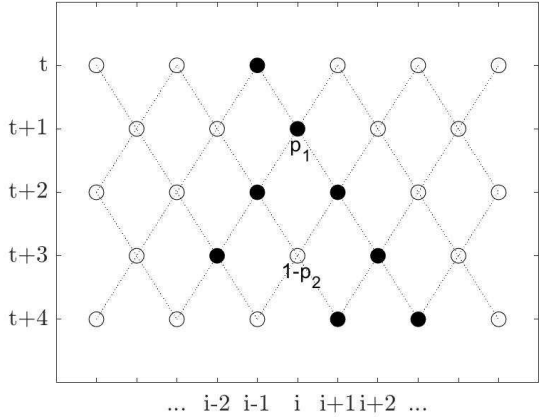}
    \caption{The $1+1$ DK model. Binary-valued sites in a tilted square grid are either active ($\bullet$) or inactive ($\circ$). Examples of its transition probabilities as defined in~\eqref{eq_DK_transition}: 1. $s_{i,t+1}=1$ with probability $p_{1}$ given its adjacent neighbors at time $t$ are in opposite states. 2. $s_{i,t+3}=0$ with probability $1-p_{2}$ when both preceding neighbors are active.}
    \label{fig_DK}
\end{figure}

A cellular automaton consists of a regular grid of discrete finite-state sites evolving according to some pre-defined parallel update rule with a discrete time variable $t\in\mathbb{N}$. The celebrated Domany-Kinzel (DK) model~\cite{Domany1984} is a stochastic cellular automaton defined on a tilted square lattice\footnote{Commonly known also as the $1+1$ DK model as referring to a single spatial dimension plus a single perpendicular temporal dimension composing together a two-dimensional square lattice.}, as depicted in Fig.~\ref{fig_DK}. The $i$'th site at time instance $t$, $s_{i,t}$, in the square lattice can be either active (occupied), hence $s_{i,t}=1$, or inactive (empty), thus $s_{i,t}=0$.

Referring to the notations in Fig.~\ref{fig_DK}, the DK model evolves stochastically following the conditional transition probabilities $\Pr(s_{i,t+1}| s_{i-1,t},s_{i+1,t})$ which depend on two parameters $p_{1}, p_{2}\in[0,1]$, where
\BEA\label{eq_DK_transition}
    \Pr(s_{i,t+1}=1|s_{i-1,t}=0,s_{i+1,t}=0)&=&0,\nonumber\\
    \Pr(s_{i,t+1}=1|s_{i-1,t}=1,s_{i+1,t}=0)&=&p_{1},\nonumber\\
    \Pr(s_{i,t+1}=1|s_{i-1,t}=0,s_{i+1,t}=1)&=&p_{1},\nonumber\\
    \Pr(s_{i,t+1}=1|s_{i-1,t}=1,s_{i+1,t}=1)&=&p_{2},
\EEA
and also evidently
\BE
    \Pr(s_{i,t+1}=0|s_{i-1,t},s_{i+1,t})=1-\Pr(s_{i,t+1}=1|s_{i-1,t},s_{i+1,t}).
\EE
This means a site at time $t+1$ in the square grid becomes active either with probability $p_{2}$ if its two nearest neighbors at time $t$ are both active or with probability $p_{1}$ if only one of them is active.

\begin{figure}[t!]
    \centering
    \includegraphics[width=\columnwidth,bb=0 0 225 172]{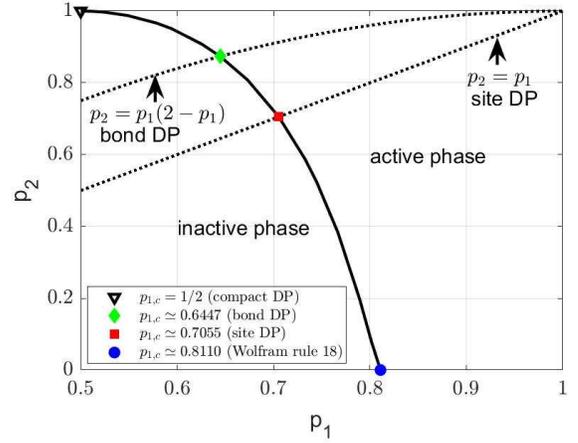}
    \caption{$1+1$ DK model phase diagram. The numerical values of the critical percolation thresholds $(p_{1,c},p_{2,c})$, composing the solid transition line, are taken from~\cite{Lubeck2006,Henkel2008}.}
    \label{fig_DK_phase}
\end{figure}

In the DK model, and in directed percolation in general, there are two main merits of interest: (a) The density, $\varrho$, of active sites in the stationary state of an infinitely large lattice.\footnote{Analogously, $\varrho$  is also the probability that a given site belongs to an infinite connected cluster that evolved from a fully occupied lattice, $s_{i,-\infty}=1, \forall i$.} (b) The percolation probability, $P_{\text{perc}}$, that a single active site in an empty lattice generates an infinite cluster successfully percolating to $t\rightarrow\infty$.

As expressed in its underlying transition probabilities~\eqref{eq_DK_transition}, the DK model is essentially controlled by the two parameters $p_{1}$ and $p_{2}$ and gives rise to a two-dimensional phase diagram, as shown in Fig.~\ref{fig_DK_phase}.  The active ($0<\varrho,P_{\text{perc}}\leq1$) and inactive ($\varrho=P_{\text{perc}}=0$) phases of the DK model are separated by a critical line of phase transition points, $(p_{1,c}=p_{c},p_{2,c})$, where $p_{c}$ is known as the percolation threshold.

Although in general the two quantities (a.k.a. order parameters) of density of active sites, $\varrho$, and percolation probability, $P_{\text{perc}}$, are different, in DP at the vicinity of the critical threshold line, $p_{1}-p_{c}\rightarrow0^{+}$, both scale algebraically with the \emph{same} critical exponent $\beta$, that is
\BE\label{eq_beta}
    \varrho\propto(p_{1}-p_{c})^{\beta},\quad P_{\text{perc}}\propto(p_{1}-p_{c})^{\beta}.
\EE
Moreover, for the particular case of bond DP, as discussed in the sequel and in Section~\ref{sec_bDP}, these two order parameters also have the same scaling factor, and furthermore $\varrho=P_{\text{perc}}$, thus can be used interchangeably.

Note that there is a strong numerical evidence in the percolation theory literature that the critical behaviour along the entire phase transition line (except for its upper end point $\triangledown$), as plotted in~Fig.~\ref{fig_DK_phase}, is the same, exhibiting a second-order\footnote{That is $\varrho$ and $P_{\text{perc}}$ are continuous, but their derivatives are discontinuous.} phase transition. This is what is known in percolation theory as the \emph{directed percolation universality class} for which the critical exponent $\beta$~\eqref{eq_beta} is identical.
Now, as mentioned, there is one exceptional point at $(p_{1,c}=1/2,p_{2,c}=1)$, known as compact DP~\cite{Domany1984,Henkel2008}, and although its misleading name it is not part of the DP universality class as it exhibits a first-order phase transition, meaning the density $\varrho$ itself is discontinuous with a sharp jump from 0 to 1 at $p_{1,c}=1/2$, thus in this case $\beta=0$.\footnote{As a matter of fact, in compact DP the percolation probability, $P_{\text{perc}}$, no longer scales similarly to the density, $\varrho$, and an additional critical exponent $\beta'=1$ must be introduced.}

The DK model includes, as special cases, the classical \emph{bond and site DP}. In bond DP, the site $s_{i,t+1}$ is activated with probability $p_{1}=p\in[0,1]$ if only one of its nearest neighbors was active at time $t$, while it is activated with probability $p_{2}=1-(1-p)^{2}=2p-p^{2}$ if both of them were active. Similarly in site DP, sites instead of bonds, are permeable with probability $p$ and blocked otherwise, thus $p_{1}=p_{2}=p$. The percolation threshold of bond ({\color{green}$\blacklozenge$}) and site ({\color{red}$\blacksquare$}) DP are denoted in Fig.~\ref{fig_DK_phase} at the intersection of these (dotted) lines, expressing the above stated relations between $p_{1}$ and $p_{2}$, with the phase transition (solid) line. The special case of $p_{2}=0$ ({\color{blue}$\bullet$}) corresponds to what is known as rule `W18' of Wolfram's code of (deterministic) cellular automata~\cite{Wolfram1983}.

Now, by definition, the average density of active sites at row $t$ is, $\varrho_{t}\triangleq\bbE(s_{i,t})$, where $\bbE(\cdot)$ denotes the expectation on the binary value of an arbitrary site, $i$, at the (stationary) state $t$. Then one gets
\BEA
    \varrho_{t+1} &=& \Pr(s_{i,t+1}=1)\nonumber\\&=&\Pr(s_{i,t+1}=1| s_{i-1,t}=1,s_{i+1,t}=0)\times\nonumber\\&&\Pr(s_{i-1,t}=1,s_{i+1,t}=0)+\nonumber\\&&\Pr(s_{i,t+1}=1| s_{i-1,t}=0,s_{i+1,t}=1)\times\nonumber\\&&\Pr(s_{i-1,t}=0,s_{i+1,t}=1)+\nonumber\\&&\Pr(s_{i,t+1}=1| s_{i-1,t}=1,s_{i+1,t}=1)\times\nonumber\\&&\Pr(s_{i-1,t}=1,s_{i+1,t}=1)+\nonumber\\&&\Pr(s_{i,t+1}=1| s_{i-1,t}=0,s_{i+1,t}=0)\times\nonumber\\&&\Pr(s_{i-1,t}=0,s_{i+1,t}=0).
\EEA
Hence incorporating the conditional transition probabilities of the $1+1$ DK model~\eqref{eq_DK_transition}, the density can be rewritten as
\BEA
    \varrho_{t+1} &=& \Pr(s_{i,t+1}=1)\nonumber\\&=& p_{1}\Pr(s_{i-1,t}=1,s_{i+1,t}=0)+\nonumber\\&&p_{1}\Pr(s_{i-1,t}=0,s_{i+1,t}=1)+\nonumber\\&&
    p_{2}\Pr(s_{i-1,t}=1,s_{i+1,t}=1).
\EEA

Now let us pick a working point $p_{1,c}^{*}=p_{c}^{*}=(1/2+\epsilon)$ (where $\epsilon\rightarrow0^{+}$) on the critical threshold line. This point is at the vicinity of the compact DP point, but resides within the DP universality class. Avoiding in our analysis the compact DP point, for which $\beta=0$, resonates with our aim to circumvent the trivial recursion, as previously stated in Footnote~\ref{foonote_trivial}. Consequently, the recursion rule for the probability of the site occupancy, in the neighborhood of $p_{c}^{*}$ in the active phase, approximately adheres to
\BEA\label{eq_percolation_rule}
     \Pr(s_{i,t+1}=1)&\simeq&\frac{1}{2}\big(\Pr(s_{i-1,t}=1,s_{i+1,t}=0)+\nonumber\\&&\Pr(s_{i-1,t}=1,s_{i+1,t}=1)\big)\nonumber\\
     &&+\nonumber\\&&
    \frac{1}{2}\big(\Pr(s_{i-1,t}=0,s_{i+1,t}=1)+\nonumber\\&&\Pr(s_{i-1,t}=1,s_{i+1,t}=1)\big)\\&=&\frac{\Pr(s_{i-1,t}=1)+\Pr(s_{i+1,t}=1)}{2}\nonumber.
\EEA

Note the resemblance of the above percolation-related recursion rule on $\Pr(s_{i,t+1}=1)$~\eqref{eq_percolation_rule} with the approximation of the polarization recursion rule of $p_{n+1}(z,c,d)$ \eqref{eq_p_rec_approx} (with $t=n$). Since there are $2^{t+1}-1\xrightarrow{t\rightarrow\infty}2^{t}$ nodes (sites) in the original binary tree, then the gap from the percolation probability, $p_{1}=p$, to the threshold, $p_{c}$, can not decay to zero faster than exponentially with $t$, namely $p-p_{c}\propto2^{-t}$. Substituting this diminishing gap into the scaling law of the DP universality class~\eqref{eq_beta}, yields
\BEA\label{eq_limiting_value_density}
    \beta&=&-\lim_{t\rightarrow\infty}\frac{1}{t}\log_{2}(\varrho)=-\lim_{t\rightarrow\infty}\frac{1}{t}\log_{2}(P_{\text{perc}})\nonumber\\&=&-\lim_{t\rightarrow\infty}\frac{1}{t}\log_{2}\big(\Pr(s_{i,t}=1)\big).
\EEA
Due to the observed similarity in the behavior of percolation's \mbox{$\Pr(s_{i,t}=1)$} and polarization's \mbox{$p_{n}(z,c,d)$}, comparing~\eqref{eq_limiting_value_density} to the limiting rate~\eqref{eq_limiting_value}, one can conclude to this end, that
\BE\label{eq_beta_mu_relation}
    \beta\simeq\tilde{\mu}^{-1}=\mu^{-1}.
\EE
Bear in mind again that the approximation, rather than equality, in~\eqref{eq_beta_mu_relation} stems from approximating the polarization tree via a percolating square grid.
Hence in order to evaluate the polar code's scaling exponent, $\mu$, one can alternatively infer the percolation critical exponent, $\beta$.

Based on the rich literature of DP and past extensive numerical investigations of the critical exponent, $\beta$, one can already provide a numerical approximation for the scaling exponent from the viewpoint of the proposed DP approach. The best known numerical estimate of $\beta$ in DP literature is given in~\cite{Jensen1999}, to yield
\BEA\label{eq_beta_numeric}
    \mu_{\text{num}}^{\text{perc}}\simeq\beta_{\text{num}}^{-1}\simeq\Big(0.276486(8)\Big)^{-1}\simeq&3.617.
\EEA

Remarkably, relying on the universality of the scaling laws of DP, one can choose any working point on the phase transition line (plotted in Fig.~\ref{fig_DK_phase}, and as discussed excluding the compact DP case). In the following section we choose to concentrate in bond DP, as a particular instantiation of the DK model, which lends itself to the derivation of a closed-form approximation for the scaling exponent, $\mu$, adopting the analytic route taken in~\cite{Sire2002}.

\subsection{Bond Directed Percolation}\label{sec_bDP}
In bond DP, bonds in the $1+1$ square grid (Fig.~\ref{fig_DK}) are randomly eliminated with probability $1-p$ (\ie, in the DK model's terminology $p_{1}=p, p_{2}=2p-p^{2}$). Note again that in bond DP the stationary site density, $\varrho$, which is the probability in steady state that a site is active ($s_{i,t}=1$), is identical to the percolation probability, $P_{\text{perc}}$, that at least one site at time $t\rightarrow\infty$ is connected to a single originating site at time $t=0$. In this section we denote these identical probabilities as $p_{i,t}$.

Thus knowing the configuration of active sites at time $t$, one can compute the next configuration at time $t+1$ according to the following straightforward probabilistic rule of bond DP
\BE\label{eq_bDP}
    p_{i,t+1}=p\cdot p_{i-1,t}+p\cdot p_{i+1,t}-p^{2}\cdot p_{i-1,t}p_{i+1,t}.
\EE
Taking expectation with both sides of~\eqref{eq_bDP}, we get in the stationary limit
\BE\label{eq_density}
    \varrho(p) = \bbE(s_{i,t}) = 2p\bbE(s_{i,t})-p^{2}\bbE(s_{i-1,t}s_{i+1,t}).
\EE

Assuming a mean-field (MF) approximation, meaning one only accounts for the single-site probabilities and simply factorizes the pairwise correlation, namely
\BE
    \bbE(s_{i-1,t}s_{i+1,t})=\bbE(s_{i,t})^{2},
\EE
then the density~\eqref{eq_density} in its MF approximation adheres to the closed-form expression
\BE\label{eq_density_MF}
    \varrho_{\text{MF}}(p) = \frac{2p-1}{p^{2}}.
\EE
From~\eqref{eq_density} we can also rewrite the exact active site density in terms of its MF approximation
\BE\label{eq_prod}
    \varrho_{\text{MF}}(p)\varrho(p) = \bbE(s_{i-1,t}s_{i+1,t}).
\EE

The average density and pairwise correlation can be written in terms of their corresponding respective derivatives, denoted as $\rho$ and $\rho_{2}$, where
\BEA
    \varrho(p)&\triangleq&\int_{0}^{p}\rho(x)dx,\label{eq_derivative}\\
    \bbE(s_{i-1,t}s_{i+1,t})&\triangleq&\int_{0}^{p}\int_{0}^{p}\rho_{2}(x_{1},x_{2})dx_{1}dx_{2}\label{eq_derivative_2}.
\EEA

It is well-known that for bond DP the percolation threshold is about $p_{1,c}=p_{c}\simeq0.6447$~\cite{Henkel2008}. As can be seen from~\eqref{eq_density_MF}, for $p>p_{c}$ the MF density resides within the $\mathcal{O}(1)$ range $[\simeq0.6963, 1]$. Hence one can approximate the logarithm of the left hand side of~\eqref{eq_prod}, using~\eqref{eq_derivative}, as
\BEA\label{eq_rhs}
    \log\big(\varrho_{\text{MF}}(p)\varrho(p)\big)&=&\log\big(\varrho_{\text{MF}}(p)\int_{0}^{p}\rho(x)dx\big)\nonumber\\
    &=&\log\big(\varrho_{\text{MF}}(p)-\int_{p}^{1}\varrho_{\text{MF}}(p)\rho(x)dx\big)\nonumber\\&\simeq&-\int_{p}^{1}\varrho_{\text{MF}}(p)\rho(x)dx.
\EEA
Similarly, taking the logarithm of the right hand side of ~\eqref{eq_prod}, this time with~\eqref{eq_derivative_2}, we also get
\BEA\label{eq_lhs}
    \log\big(\varrho_{\text{MF}}(p)\varrho(p))&=&\log\int_{0}^{p}\int_{0}^{p}\rho_{2}(x_{1},x_{2})dx_{1}dx_{2}\nonumber\\
    &\simeq&-(1-\int_{0}^{p}\int_{0}^{p}\rho_{2}(x_{1},x_{2})dx_{1}dx_{2})\nonumber\\
    &=&-\int_{0}^{p}\int_{p}^{1}\rho_{2}(x_{1},x_{2})dx_{1}dx_{2}\nonumber\\&-&\int_{p}^{1}\int_{0}^{1}\rho_{2}(x_{1},x_{2})dx_{1}dx_{2}.
\EEA

Writing the pairwise correlation's stationary probability distribution in a form supporting a non-zero probability that $x_{1}=x_{2}$, one gets
\BE\label{eq_formulation}
    \rho_{2}(x_{1},x_{2})=\tilde{\rho}(x_{1},x_{2})+\rho(x_{1})g(x_{1})\delta(x_{1}-x_{2}),
\EE
where $g(x)$ is defined as the probability that $x_{2}=p$, conditional to its neighbor being $x_{1}=p$, and $\tilde{\rho}(x_{1},x_{2})$ is the corresponding two-point probability density function without a point-mass. $\delta(\cdot)$ is the Dirac delta function. Now comparing~\eqref{eq_rhs} and~\eqref{eq_lhs}, utilizing the formulation ~\eqref{eq_formulation}, immediately yields the approximate relation
\BE\label{eq_almost_final}
    \varrho_{\text{MF}}(x)\rho(x)\simeq\rho_{\text{MF}}(x)+\rho(x)g(x),
    %\varrho_{\text{MF}}(p)\rho(x) = \rho_{\text{MF}}(x)+\rho(x)g(x),
\EE
where
\BE
    \rho_{\text{MF}}(x)\triangleq\frac{d\varrho_{\text{MF}}(x)}{dx}=\tilde{\rho}(x)=\int_{0}^{1}\tilde{\rho}(x,x_{2})dx_{2}.
\EE
Rearranging~\eqref{eq_almost_final}, we get that
\BE\label{eq_almost_final2}
    \rho(x)\simeq\frac{\rho_{\text{MF}}(x)}{\varrho_{\text{MF}}(x)-g(x)}.
    %\rho(x)=\frac{\rho_{\text{MF}}(x)}{\varrho_{\text{MF}}(p)-g(x)}.
\EE

Now, the density (or interchangeably the percolation probability) can be approximated as
\BEA\label{eq_varrho}
    \varrho(p)&=&\exp\Big(\log\big(\varrho(p)\big)\Big)\nonumber\\&=&\exp\Big(\log\big(\int_{0}^{p}\rho(x)dx\big)\Big)\nonumber\\
    &=&\exp\Big(\log\big(1-\int_{p}^{1}\rho(x)dx\big)\Big)\nonumber\\
    &\simeq&\exp\Big(-\int_{p}^{1}\rho(x)dx\Big)\nonumber\\
    &\simeq&\exp\Big(-\int_{p}^{1}\frac{\rho_{\text{MF}}(x)}{\varrho_{\text{MF}}(x)-g(x)}dx\Big)\label{eq_last_equality}\\
    %&=&\exp\Big(-\int_{p}^{1}\frac{\rho_{\text{MF}}(x)}{\varrho_{\text{MF}}(p)-g(x)}dx\Big)\label{eq_last_equality}\\
    &\triangleq&\exp\Big(-\int_{p}^{1}\frac{f_{1}(x)}{f_{2}(x)}dx\Big)\label{eq_last_equality2},
\EEA
where in~\eqref{eq_last_equality} we have applied the approximate relation~\eqref{eq_almost_final2}.

From~\eqref{eq_beta} we know that as $p\rightarrow p_{c}$, the density $\varrho$ vanishes according to $(p-p_{c})^{\beta}$. Hence the function within the integral in~\eqref{eq_last_equality2}, must have a single pole at $p_{c}$ of residue $\beta$. This means that $f_{2}(p_{c})=0$, thus $\varrho_{\text{MF}}(p_{c})=g(p_{c})$, and the residue, which is the critical exponent, can be obtained from the L'Hôpital's rule as
\BE\label{eq_compute_beta}
    \beta \simeq \frac{f_{1}(x)}{df_{2}(x)/dx}\Big{|}_{x=p_{c}}=\frac{\rho_{\text{MF}}(p_{c})}{\rho_{\text{MF}}(p_{c})-dg(x)/dx|_{x=p_{c}}}.
\EE

We have already computed~$\varrho_{\text{MF}}(p)$ explicitly in~\eqref{eq_density_MF}. Thus taking its derivative w.r.t. $p$, we find
\BE\label{eq_rho_MF}
    \rho_{\text{MF}}(p)=\frac{d\varrho_{\text{MF}}(p)}{dp}=\frac{2(1-p)}{p^{3}}.
\EE
Hence in order to compute $\beta$ from~\eqref{eq_compute_beta} we are only missing the expression for $g(p)$.

Here is where the \emph{golden ratio} comes into action. We first observe that in bond DP the percolation threshold
\BE\label{eq_crude_approximation}
    p_{c}\simeq0.6447\approx\Phi\simeq0.618,
\EE
where
\BE
    \Phi\triangleq1/\varphi=\varphi-1
\EE
is known as the \emph{golden ratio conjugate}. We will see how the relatively crude approximation of the percolation threshold of bond DP as the golden ratio conjugate in~\eqref{eq_crude_approximation}, $p_{c}\approx\Phi$, entails a remarkable prediction of its critical exponent $\beta$.

Now, note that
\BE
    \varrho_{\text{MF}}(\Phi)=\frac{2\Phi-1}{\Phi^{2}}=\Phi=g(\Phi).
\EE
Hence, for two scalars $a$ and $b$ with $a>b$ for which
\BE
    \varrho_{\text{MF}}(\Phi)\triangleq\frac{b}{a}=\Phi,
\EE
then since these two scalars \emph{are in the golden ratio}, also
\BE\label{eq_golden_relation}
    g(\Phi)=\frac{a-b}{b}=\Phi,
\EE
which immediately yields the missing explicit expression
\BE
    g(p)=\frac{(1-p)^{2}}{2p-1}.
\EE

Thus now one can also easily compute the derivative $dg(p)/dp|_{p=\Phi}$, along with $\rho_{\text{MF}}(\Phi)$ from~\eqref{eq_rho_MF}, and substitute them into~\eqref{eq_compute_beta} to get\footnote{In~\eqref{eq_golden_relation} one can alternatively use the golden relation $g(\Phi)=a/(a+b)=\Phi$, to infer a different expression for $g(p)$. However, substituting the derivative of this second solution, at $p=\Phi$, in \eqref{eq_compute_beta} interestingly yields the complementary value $\beta_{2}=1-\beta$, which is not a valid solution for the scaling exponent as resulting in $\mu<2$. Thus the golden relation~\eqref{eq_golden_relation} was chosen.}
\BE
    \beta \simeq (2+1+\Phi)^{-1}=(2+\varphi)^{-1}\simeq0.276393.
\EE
Note how close (within only 0.028\%) is this approximation to the best known numerically computed estimate $\beta_{\text{num}}=0.276486(8)$ from~\cite{Jensen1999}.
Using the relation~\eqref{eq_beta_mu_relation}, we can finally conclude
\BE
    \mu=\tilde{\mu}\simeq\beta^{-1} \simeq 2+\varphi\simeq3.618,
\EE
which proves the main result~\eqref{eq_main_result}.

Recapping on the analysis, it consists of two approximation insights: (a) First, an approximation of the polarization on the binary tree to directed percolation in a $1+1$ DK model. (b) And then applying another approximation based on the golden number relations to facilitate an analytical, rather than numerical, expression for the $1+1$ DK model's critical exponent, $\beta$. Thus it is important to note that the approximation error in~\eqref{eq_main_result} mainly stems from the former, rather than the latter, approximation concept.

\section{Conclusion}
The yet unknown, in closed-form, penalty in the scaling exponent of the conventional polar code in BEC, w.r.t. the optimal value of 2, achieved by random coding, is analytically approximated by the golden ratio, which ubiquitously appears across the history of science, design and art. The derived scaling exponent, born from pure percolation analysis, falls within the previously known tight analytical bounds and beautifully approximates numerical estimate.

From a wider viewpoint, to the best of our knowledge this is the first successful attempt in applying (directed) percolation theory as a tool for analyzing ECCs, showing that the universal scaling laws of DP can be effectively utilized in the study of the intrinsic characteristics of polar codes. We hope it may help in igniting a proliferation of valuable results in the interface of percolation theory and coding.

%, similar to the effect of Sourlas’ 1989 Nature letter~\cite{Sourlas1989} leading the way for a plethora of literature on the interface of %spin-glass theory and coding.

\section*{Acknowledgment}
The author thanks Wolfgang Kinzel and Hamed Hassani for their useful comments on an earlier version of this paper.

\bibliographystyle{IEEEtran}
\bibliography{IEEEabrv,golden_polar_Ori_19}

\end{document}